\documentclass{nature}
\usepackage{epsfig}

\usepackage{amsmath}

\begin{document}

\title{State-independent experimental test of quantum contextuality in solid state system}
\author{Xin-Yu Pan$^\star$, Yan-Chun Chang, Gang-Qin Liu, and Heng Fan$^\star$}

\maketitle

\begin{affiliations}
\item
Beijing National Laboratory for Condensed Matter Physics, Institute
of Physics, Chinese Academy of Sciences, Beijing 100190, China
\\

$^\star$e-mail: xypan@aphy.iphy.ac.cn; hfan@iphy.ac.cn;

\end{affiliations}

\begin{abstract}
Quantum mechanics implies that not all physical properties can be simultaneously well defined,
such as the momentum and position due to Heisenberg uncertainty principle. Some alternative
theories have been explored, notably the non-contextual hidden variable theories in which the
properties of a system have pre-defined values which are independent of the measurement contextual.
However, the Kochen-Specker theorem \cite{Kochen-Specker} showed that such non-contextual
hidden variable theories are in conflict with quantum mechanics. Recently, a state-independent inequality
satisfied by non-contextual hidden variable theories and violated by quantum mechanics is proposed
in the simplest three-state system (a qutrit)
by the least 13 projection measurement rays \cite{YuOh}.
Here, we report an experimental demonstration of the violation of this
inequality. This provides
a state-independent experimental test of quantum contextuality,
for the first time in solid state
system, by a natural qutrit of nitrogen-vacancy center in diamond at room temperature.
\end{abstract}

The question of whether quantum phenomena can be explained by non-contextual hidden variable theories
has a long history \cite{EPR,Mermin}. Non-contextuality means that the measured value of an observable is independent
of its own measurement and other co-measurable observables that are measured previously or simultaneously.
This is in consistent with our experience in everyday lives.
However, as proved by Kochen and Specker \cite{Kochen-Specker}, and also by Bell \cite{Bell}, the non-contextual hidden variable
theories cannot reproduce quantum mechanics because quantum theory predicts that the outcomes
depend on the context of measurement. This is proven either by the violation of Bell type inequalities since
of the quantum nonlocality or by a logical contradiction between the local hidden variable
predictions and those of quantum mechanics. The violation of Bell type inequalities depends on
specified quantum entanglement state. The state-independent Kochen-Specker theorem is proven by a logical contradiction
which applies to systems with dimension of Hilbert space larger than three.

The proof of Kochen-Specker theorem involves initially 117 observables in a qutrit system which is the simplest
system capable of manifesting the quantum contextuality \cite{Kochen-Specker}. The number of observables involved in the proof
is reduced gradually to 31 rays for state-independent case \cite{Peres,Conway-Kochen} and
5 rays for state-dependent case \cite{Klyachko}.
Recently, Yu and Oh \cite{YuOh} provide a state-independent proof with only 13 rays for a qutrit,
which is optimal \cite{Cabello}.
The contradiction in this proof involves a Yu-Oh inequality which is consistent with non-contextual hidden
variable theories but should be violated by any quantum states predicted by quantum mechanics.
Additionally, the system only involves a single
qutrit which is dimension three and is thus indivisible, so the contradiction cannot result from entanglement.

The quantum contextuality is continuously tested experimentally in two quantum bits (qubits) system by using photons \cite{Huang},
neutrons \cite{Hasegawa,Bartosik}, ions \cite{Kirchmair}, ensembles of nuclear magnetic resonance \cite{Moussa}.
In a qutrit system, the state-dependent \cite{Klyachko} experimental test is performed by single photons \cite{Lapkiewicz}.
The newly proposed Yu-Oh scheme is just realized experimentally also by photons \cite{Zu}.
In this work, we report an experimental test of quantum contextuality, for the first time in solid-state system,
by nitrogen-vacancy (NV) center in diamond at room temperature.
Our experimental results strongly agree with quantum contextuality manifested in a state-independent characteristic.

A NV center comprises a substitutional nitrogen atom instead of a
carbon atom and an adjacent lattice vacancy in diamond, which provides an
electronic spin \cite{Gruber97,Jelezko1}, see Fig.1.
The electronic spin of NV
center in diamond can be individually addressed, optically
polarized, manipulated and measured with
optical and microwave excitation.
Since those properties and its long coherence time, the NV center system stands
out as one of the most promising solid state systems as quantum
information processors \cite{Gaebelwrachtrupnaturephysics06,Childressscience,
Jianglukinscience09,Neumannnaturephysics10,Shiduprl10}
and as a high-sensitivity magnetometer \cite{Mazelukinnature,Taylornaturephys}.

The electronic ground state of the NV center is a spin triplet,
and it exhibits a zero field splitting defining the
$\hat {z}$ axis of the electron spin.
The state with zero magnetic moment $m_s=0$ is corresponding to $|0\rangle $,
the application of a magnetic
field splits the magnetic sub-levels $m_s=\pm 1$
corresponding to $|\pm 1\rangle $, respectively, which allow selective
microwave excitation.
A qubit of NV center, which involves $m_s=0$ and a single spin transition
to either $m_s=1$ or $m_s=-1$, has been widely demonstrated.
A qutrit is a superposed three
level state which will involve all three energy levels of the electronic spin in a NV center.
The precise control of
an arbitrary qutrit in NV center is also demonstrated experimentally in a quantum information
processing scheme in our group \cite{APL}. This provides a perfect
experimental setup to test quantum contextuality of Yu-Oh scheme \cite{YuOh}, for a large variants of
quantum states to illustrate the state-independent property.

The Yu-Oh proof of quantum contextuality involves arbitrary qutrits and 13 projection
measurement rays \cite{YuOh}. For a given basis $\{ |0\rangle ,|1\rangle ,|-1\rangle \}$ which are
realized by three electronic spin energy levels of a NV center in our experiment, $m_s=0,\pm 1$,
the 13 normalized rays are represented as rank-1 qutrit projectors $\hat {r}=|r\rangle \langle r|$ in which
$|r\rangle =a|0\rangle +b|1\rangle +c|-1\rangle $ (see Methods).
Suppose $a_v$ are outcomes corresponding to those 13 observables, $(a_ua_v)$ are the correlations,
it is shown \cite{YuOh} that an inequality should be satisfied by non-contextual hidden variable theories,
even though the usual Kochen-Specker value assignment does exist for those 13 rays,
\begin{eqnarray}
\sum _ua_u-\frac {1}{4}\sum _{\langle u,v\rangle }(a_ua_v)\le 8,
\label{YuOh-inequality}
\end{eqnarray}
where $\langle u,v\rangle $ denotes all 24 compatible pairs of observables (see Fig.2 and Methods).
On the other hand, this inequality
should be violated by quantum mechanics since it is expected by quantum theory
that $\sum _ua_u-\frac {1}{4}\sum _{\langle u,v\rangle }(a_ua_v)= 25/3\approx 8.33$, no matter what kind of
quantum states are measured.

In NV center system, the experimental test of quantum contextuality
is to prepare arbitrary qutrits, measure 13 observables and 24 correlations for each qutrit.
The aim is to find whether the inequality (\ref{YuOh-inequality}) is violated or not in a state-independent nature.
The experiment is carried out in type IIa sample diamond
(nitrogen concentration $\ll 1 $ ppm ),
most of the atoms (12C, natural abundance of $98.9\% $) have no nuclear spin, result in a very pure spin bath and thus the
NV electron spin has long enough coherence time for manipulation and readout of the qutrit.
But due to the hyperfine interaction with the host nitrogen nuclear spin, each of the electron spin state splits into three sub-levels,
and it is not sufficient to precise manipulate all the three sub-states by a single pulse.
Fortunately, we can polarize the nuclear spin by adding an external magnetic field of about 500 Gauss along NV symmetry axis \cite{add}.
This magnetic field also lifts the degeneration between $m_s=\pm 1$ states.
In Fig.1(c) we plot the optically detected magnetic resonance (ODMR) spectral of this center. The sharp dip at frequency of 1480.6 MHz and 4259.3 MHz are corresponding to the $|0> \leftrightarrow |-1>$ and $|0>\leftrightarrow |1>$ transitions. Fig.1(d) presents typical Rabi oscillation of this center. The perfect oscillation reveals that the microwave pulse can control the electron spin with high fidelity.
The 532 nm laser beam is modulated by an acousto-optic modulation (AOM) and then reflected to the objective by a galvanometer, which controls the focus position on the sample.
The fluorescence of $|0>$ state is about $30\%  $
brighter than the $|\pm 1>$ states,
so the spin state of NV center can be readout by applying a short laser (0.3 $\mu$s)
pulse and recording the excited fluorescence at the same time.


Two microwave sources are used with different frequencies: MW1 and
MW2. MW1 frequency is resonant with the transition $m_s$=0
$\leftrightarrow$ $m_s$=-1, and MW2 with $m_s$=0 $\leftrightarrow$
$m_s$=1, which can create superposed states of $|0\rangle $ with
$|-1\rangle $ or $|1\rangle $,respectively.
The electron spin is firstly polarized into
state $|0\rangle $ by laser pulse of duration 3.5 $\mu$s followed by
5 $\mu$s waiting.
Then the combination of
two microwaves, MW1 and MW2, will let us prepare an arbitrary qutrit
in superposition form by controlling the duration of MW1 and MW2,
see Methods.

Experimentally, we prepare 13 pure qutrits $|r\rangle $ corresponding to
13 projectors as presented in Methods section.
Other types of qutrits can be similarly prepared without any experimental difficulty
since they need same techniques.
With arbitrary qutrits prepared, the scheme of
measurement is the same: measure 13 observables and 24 pairs of correlations.
Experimentally, we realize each measurement of the 13 observables by a
qutrit rotation such that the projector $|r\rangle \langle r|$ can be transformed
to $|0\rangle \langle 0|$, which is realized by applying either MW1 or MW2 or both,
followed by a readout of the intensity of florescence corresponding to
the population in state $|0\rangle $ by normalization, see Fig.3 and Methods.
The readout can be made by two Rabi oscillations to fit the relative percentages
between $|0\rangle $ and $|\pm 1\rangle $, then the form of the
superposed qutrit can be found.
The scheme of measuring 24 correlations is implemented by
measuring the joint probabilities or by multiplication the outcomes of
compatible observables, see Methods.

Our experiment results are presented in Fig.4.
Each experimental data (point) in Rabi oscillation is the averaged value which
is repeated for $10^6$ times. For each prepared qutrit,
two Rabi oscillations, in which each contains 24 points,
are used to extract one outcome for one observable.
We have 13 pure states prepared,
for each input state, Fig.4(a) shows experimental values of outcomes of 13 observables,
Fig.4(b) are theoretical results. They agree well with each other. Thus a simpler inequality involving only four $h$-type
observables \cite{YuOh} is clearly violated for all input states.
Fig.4(c) shows two pure states as examples, the 13 outcomes of observables agree well
with theoretical expectation. Our results are summarized in Fig.4(d) which
includes measured correlations. Clearly for each prepared
state, the inequality (\ref{YuOh-inequality}) is violated.
All results are very close to theoretical prediction of quantum mechanics with range
8.33(-0.2,+0.14).
The reason that some results are larger than theoretical value 8.33 is that, the measured outcomes
are fitted with standard Rabi oscillations, so the error fluctuations can be on both directions.
Averaged result is 8.31 which is about $99.8\% $ of 8.33.
In conclusion, the state-independent violation of (\ref{YuOh-inequality}) is convincing.

In this work, we observe that the inequality (\ref{YuOh-inequality}) is
clearly violated thus the state-independent quantum contextuality is confirmed
experimentally in a natural qutrit of nitrogen-vacancy center of
diamond at room temperature. This real qutrit may
close the possible loophole appeared in
such as photonic systems induced by the post-selection technique.
This represents the first experimental test of quantum contextuality in solid-state
system, and provides new evidences in a most fundamental form
that quantum mechanics is contextual.

\section*{Methods}
\noindent \textbf{Yu-Oh proof of quantum contextuality and notations.}
Explicitly, the 13 rays take the forms corresponding to vectors
$(a,b,c)$ for $|r\rangle =a|0\rangle +b|1\rangle +c|-1\rangle $ as shown in Fig.2(c).
The orthogonality relationships among the 13 rays are schematically represented
in Fig.2(a). Correspondingly, we can define a set of 13 observables, $\hat {A}_v=1-2\hat {r}_v$, where
$\hat {r}_v\in \{ \hat {z}_k,\hat {y}^{\sigma }_k,\hat {h}_k,\hat {h}_0\}, \sigma =\pm ,k=1,2,3$, here
we use notation $\hat {r}=|r\rangle \langle r|$. When
$\hat {A}_u, \hat {A}_v$ are commuting, i.e., $|r_u\rangle $ and $|r_v\rangle $ are orthogonal,
they are co-measurable (compatible) observables so that their correlation is well defined in quantum
mechanics. In contrast, it is always well defined in a non-contextual hidden variable theory regardless of
whether they are compatible or not,
so that each outcome $a_v$ of observable $\hat {A}_v$ always takes values $\pm 1$.
In quantum theory for the density operator $\rho $ of an arbitrary quantum state,
the outcome of an observable is, $a_v=\langle \hat {A}_v\rangle = {\rm Tr}( \rho \hat {A}_v)$, and the correlation is,
$(a_ua_v)=\langle \hat {A}_u\hat {A}_v\rangle ={\rm Tr}( \rho \hat {A}_u\hat {A}_v)$, provided they are compatible.
Then different results are expected from non-contextual hidden variable theories and
quantum mechanics depending on whether the inequality (\ref{YuOh-inequality}) is satisfied or
violated. So the experimental test is only to check this inequality.

Additionally, a simpler inequality is also proposed which involves only four $h$-type rays
$\{\hat {h}_k \} _{j=0}^3$,
inequality $\sum \langle \hat {h}_k\rangle \le 1$ should be satisfied by non-contextual hidden variable theories,
while quantum mechanics expects that it should be violated and the
right hand side is $4/3\approx 1.33$.

\noindent \textbf{Preparation of arbitrary qutrits and readout.}
After initialization of the electron spin to $|0\rangle $, two channels of
resonant microwave pulses are applied to manipulate the spin state.
They are generated by individual source and timing controlled by individual RF switches,
and then combined together to a 16 Watt amplifier. The amplified microwave signal is then delivered to the sample by
a 20 $\mu $m copper wire which is mounted on the diamond surface.
For each of the microwave pulse, its phase $\varphi $  determines the rotation axis and the
pulse duration $t_{MW}$ determines the rotation angle. E.g. if we apply a MW1 pulse to state
$|0>$, and define $\theta =2\pi t_{MV}/T$ with $T$ being the duration of one circle
of Rabi oscillation for MW1, then the result state is
$\cos \frac {\theta }{2}|0\rangle +e^{i\varphi }\sin \frac {\theta }{2}|-1\rangle $.
This operation equals to rotate the state with $\theta $  angle around an axis in X-Y plane,
where the angle between the rotation axis and X-axis is $\varphi $.

The readout is by measuring the intensities of fluorescence by two Rabi oscillations
and fit the data to those two curves. Define $P_1$ and $P_2$ to represent the percentage of $|0\rangle $
among $\{ |0\rangle ,|1\rangle \} $ subspace and
$\{ |0\rangle ,|-1\rangle \} $  subspace.
By fitting the two Rabi curves we can extract $P_1$ and $P_2$. By the normalization condition, we get the projection result:
$P=\frac {P_1P_2}{P_1+P_2-P_1P_2}$. The projection results of 25 measurements for each state are combined together to calculate the
24 pairs of correlations. This readout method in NV center system is useful to reduce the effect of
decoherence, in particular for depolarizing noise.
Fig.3(a) shows an example, a test state $Y_3^-=(1,1,0)/\sqrt {2}$ is prepared.
Rabi oscillation of MW1 (upper, orange curve) shows that the superposed state between $|0\rangle $ and $|-1\rangle $ is
 $(|0\rangle +|-1\rangle )/\sqrt {2}$, while Rabi oscillation of MW2 (below, blue curve) shows that the percentage of $|1\rangle $
 is zero. So the form of state $Y_3^-=(1,1,0)/\sqrt {2}$ is confirmed.
 The detailed information of all states preparation and readout is presented in Fig.3(c).

In our experiment, the pulse length is determined by an additional Rabi oscillation of $|0\rangle $
 state (Fig.1(d)) which acts as a benchmark,
 and is updated in every two hours to ensure the accuracy. All states preparation and measurements
 are programming controlled and the experiment can run automatically for several days.

\noindent \textbf{Measurement of 24 correlations.} The 24 pair correlations involved in the inequality (\ref{YuOh-inequality}) can be written in the
form as $\langle \hat {A}_u\hat {A}_v\rangle =P_{(a_u=1,a_v=1)}+P_{(a_u=-1,a_v=-1)}-P_{(a_u=1,a_v=-1)}-P_{(a_u=-1,a_v=1)}$.
We construct an orthogonal set of rays $\{ |r_u\rangle ,|r_v\rangle ,|r_w\rangle \} $ which constitutes a
complete basis as shown in Fig.2,
note that ray $|r_w\rangle $ may not necessarily be within the existing 13 rays in Yu-Oh scheme.
All involved rays in this work
can be found in Fig.2(b,c).
Now, we have $P_{({r}_u=1)}+P_{({r}_v=1)}+P_{({r}_w=1)}=1$, where parameter $r_u$ is measured outcome of
observable $\hat {r}_u$, so four terms on the right hand
side of the above equation can be rewritten respectively as,
$P_{(a_u=1,a_v=1)}=P_{({r}_u=0,{r}_v=0)}=P_{({r}_w=1)}$;
$P_{(a_u=-1,a_v=-1)}=P_{({r}_u=1,{r}_v=1)}=
P_{({r}_u=1|{r}_v=1)}P_{({r}_v=1)}$; and $P_{(a_u=1,a_v=-1)}=
(1-P_{({r}_u=1|{r}_v=1)})P_{({r}_v=1)}$; $P_{(a_u=-1,a_v=1)}=
(1-P_{({r}_v=1|{r}_u=1)})P_{({r}_u=1)}$. In the last three equations, $P_{({r}_u=1|{r}_v=1)}$ means
the state $|r_v\rangle $ is measured by projective
ray $\hat {r}_u$,
the corresponding two states, $|r_u\rangle ,|r_v\rangle $, are actually orthogonal to each other thus
the measured outcome is very small.
So now the correlations are also
realized by projective measurements. Alternatively, we can measure $\langle \hat {A}_u\hat {A}_v\rangle $ in a framework of quantum theory, such that
the outcome of $\langle \hat {r}_u\hat {r}_v\rangle $ is the square root of a multiplication of
three quantities, which are expectation values of $\hat {r}_u$,$\hat {r}_v$ for the prepared qutrits,
and outcome of the projective measurement of state $|r_u\rangle $ by projector $\hat {r}_v$.
Those two methods give same result experimentally.

\begin{addendum}

\item [Acknowledgement]

This work was supported by ``973'' programs (2009CB929103, 2010CB922904)
and NSFC grants (11175248, 10974251).
We would like to thank S.X.Yu and C.H.Oh for useful discussions about
their paper \cite{YuOh}, we also would like to
 thank Zu and Zhang from Tsinghua for letting
us know their related paper in photonic system \cite{Zu} when this manuscript
was in preparation. We thank L.M.Duan for discussions.

\item [Author Contributions]
X.-Y.P., and H.F. designed the experiment. X.-Y.P. is in charge of experiment, H.F.
is in charge of theory. Y.-C.C.,G.-Q.L.,
and X.-Y.P. performed the experiment. H.F. wrote the
paper, with assistances from X.-Y. P., G.-Q.L. and Y.-C.C.
All authors conceived the research, analyzed the data and commented on the manuscript.
Y.-C.C. and G.-Q.L. contributed equally.

\item [Competing Interests]
The authors declare that they have no competing financial interests.

\item [Correspondence]
Correspondence and requests for materials should be addressed to
X.-Y.P or H.F.
\end{addendum}

\clearpage

\newpage

\bigskip

\bigskip
\textbf{Figure 1 The NV center system in diamond.}
 a. Structure of NV center, a carbon of diamond is replaced by a nitrogen, a vacancy is located in the nearby lattice site.
 b. A two-dimensional scan image of diamond sample, the bright spot (circled) is the NV center investigated.
c. Three energy levels and the ODMR spectra are given,
two sharp dips correspond to resonating frequencies.
 d. Rabi oscillations of MW1 (orange) and MW2 (blue).

\bigskip
\textbf{Figure 2 The orthogonality relationships among the 13 projection rays and
the complementary rays.}
a. 
The 13 rays are represented by vertices,
each connecting bond
between two vertices
represents that these two rays are orthogonal and the corresponding observables are compatible (co-measurable).
There are altogether 24 connecting bonds representing 24 correlations.
b. 
Starting from 13 rays, some complementary rays are presented to
form full sets of basis. Here, the entries are orthogonal with horizontal and vertical 13 rays.
For example, a complementary ray $x_1^0$ is orthogonal with both $h_0$ and $y_1^-$.
c. The forms of all involved vectors.

\bigskip

\textbf{Figure 3 Pulse sequences for states preparation and projective measurements.}
a. We prepare a test state $Y_3^-=(1,1,0)/\sqrt {2}$. 
Rabi oscillation of MW1 (upper, orange curve)
and MW2 (below, blue curve)
confirm the result.
b. The test state $Y_3^-$ is
 measured by observable ray $h_1=(-1,1,1)/\sqrt {3}$, the result is consistent with theoretical value 0.667.
 c. Pulse sequences to prepare 13 pure states.
 Each preparation is realized by a MW2 pulse (orange) followed by a MW1 pulse (blue),
 the readout is in reverse order.

\bigskip

\textbf{Figure 4 Measured results and conclusions.}
a. We have prepared 13 pure states (vertical axis),
the experimental values for 13 observables (horizontal axis) are presented.
b. In contrast, the theoretical results are presented.
c. As examples, we show the measurement results for two pure states,
$Z_3=(0,1,0), Y_3^+=(1,1,0)/\sqrt {2}$,
theoretical and experimental results are close to each other.
d. (upper) The summarized results for all prepared 13 pure states are presented.
The right hand side of the inequality (\ref{YuOh-inequality}) is calculated by experimental results,
the values range from
8.13 to 8.47, then inequality (\ref{YuOh-inequality}) is always violated
without exception.
The average value 8.31 is about $99.8\% $ of the theoretical
value 8.33 derived by quantum mechanics.
d. (down) 
A simpler inequality presented in Methods is also violated without exception.

\newpage
\begin{figure}
\begin{center}
\epsfig{file=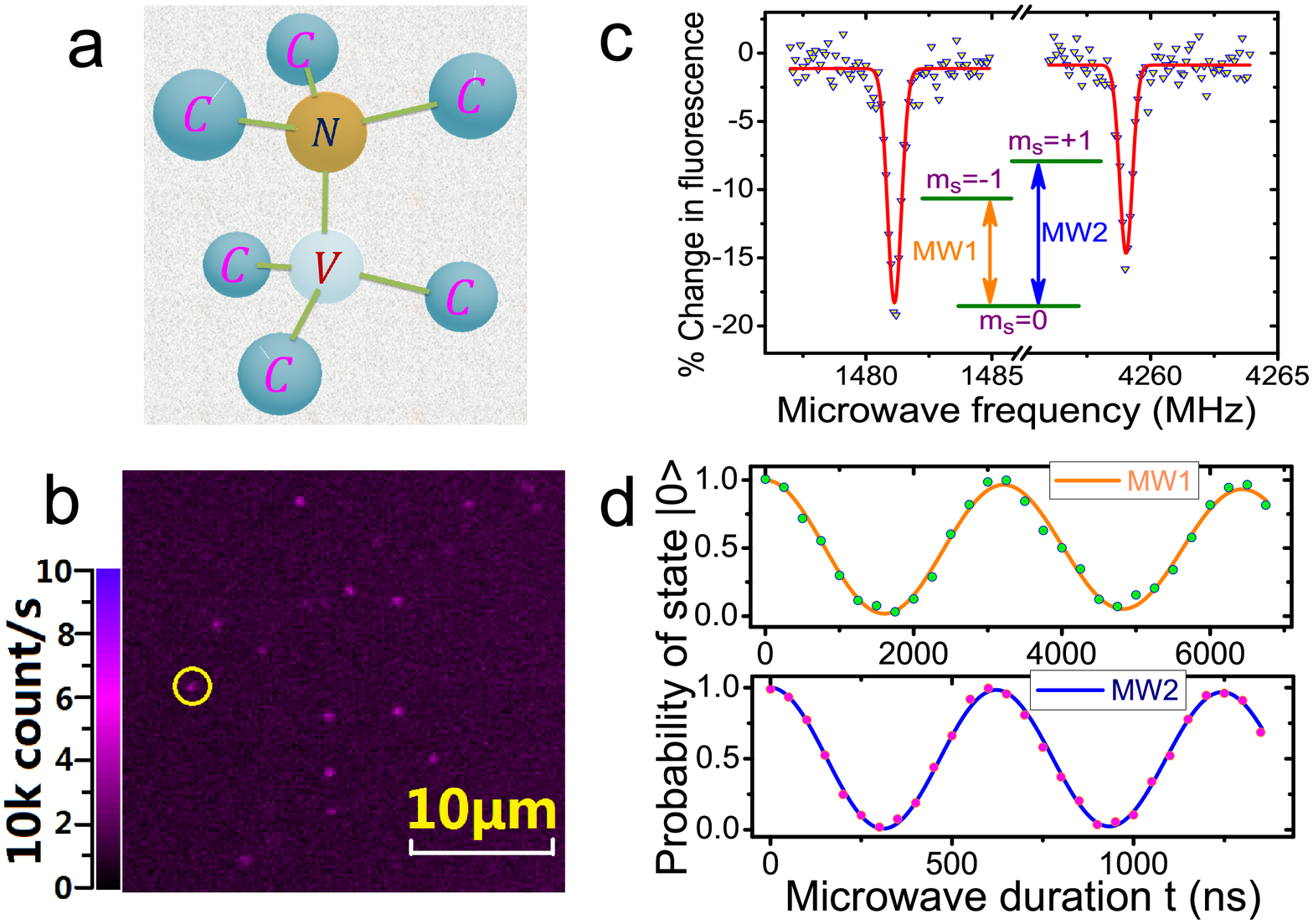,width=18cm}
\end{center}
\label{fig1}
\end{figure}

\newpage


\newpage

\begin{figure}
\begin{center}
\epsfig{file=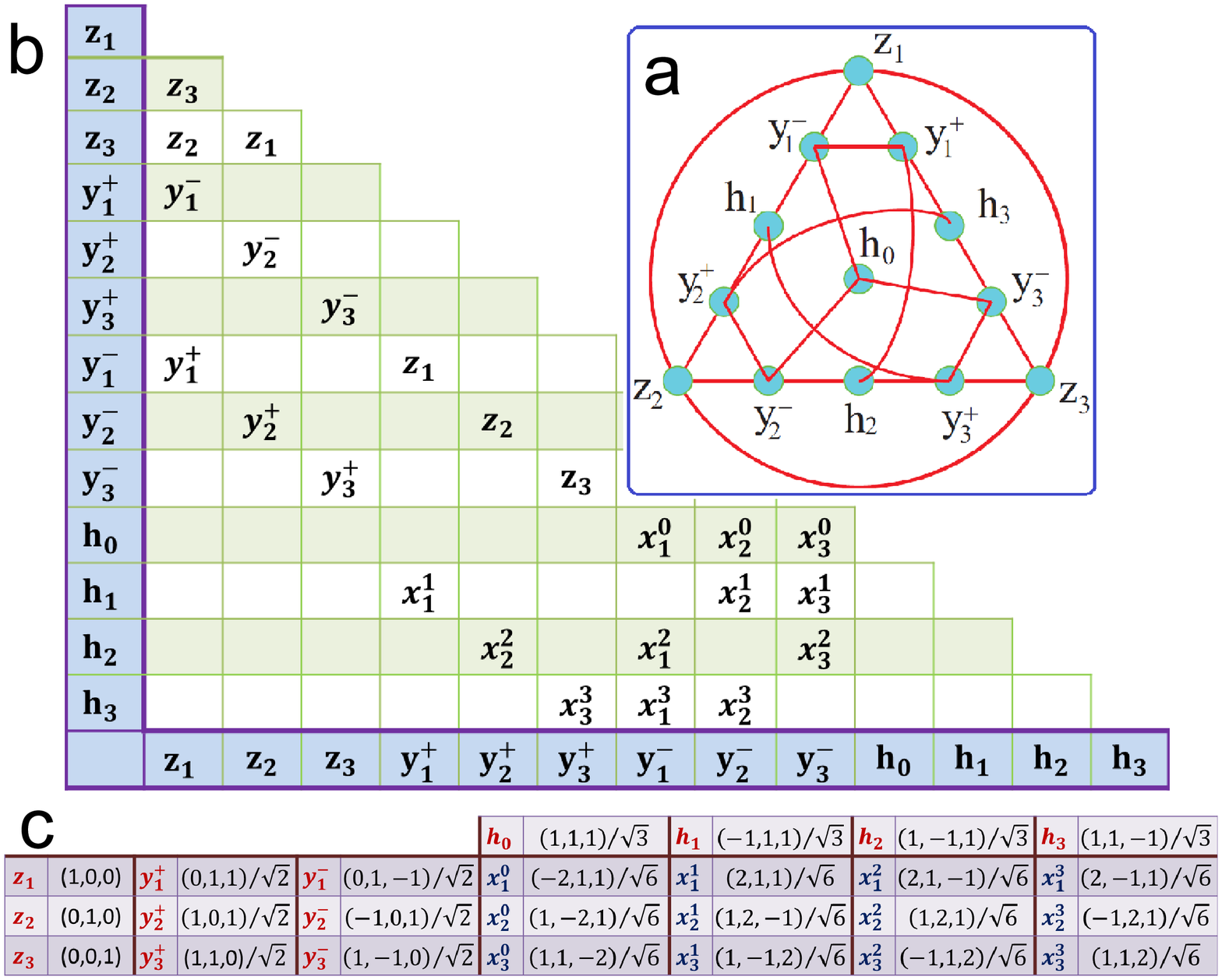,width=18cm}
\end{center}
\label{fig3}
\end{figure}

\newpage

\begin{figure}
\begin{center}
\epsfig{file=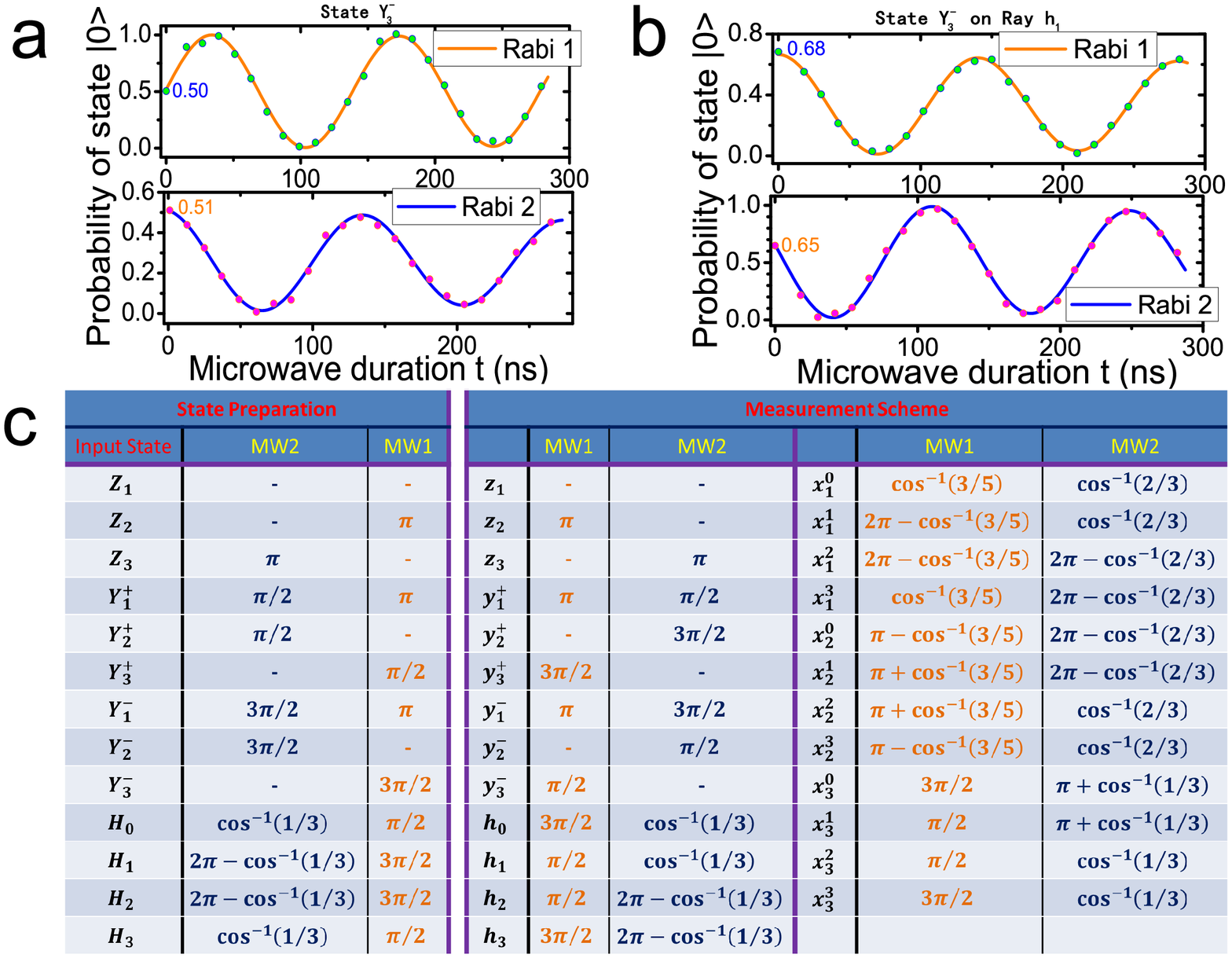,width=18cm}
\end{center}
\label{fig4}
\end{figure}

\newpage

\begin{figure}
\begin{center}
\epsfig{file=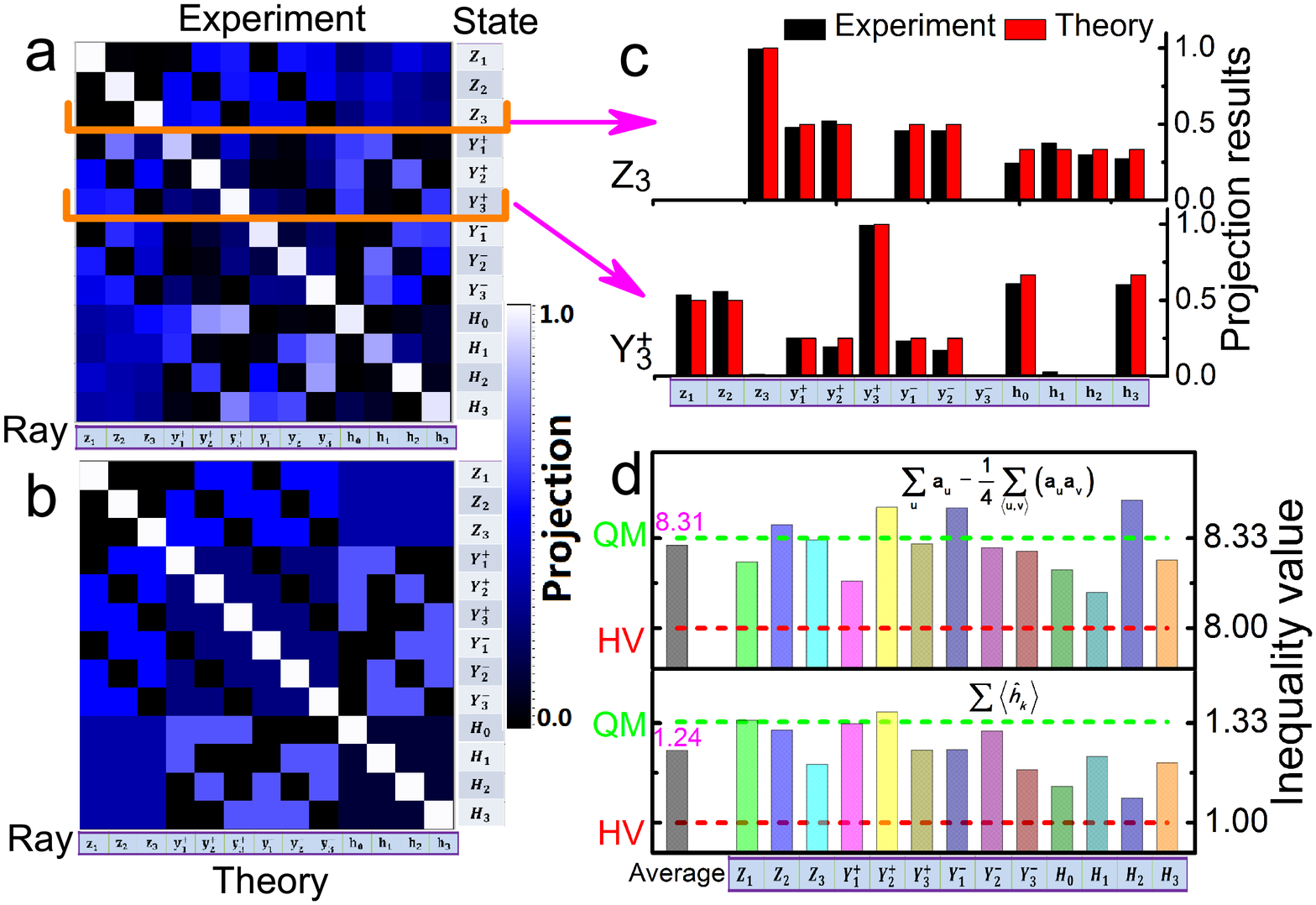,width=18cm}
\end{center}
\label{fig4}
\end{figure}

\end{document}